\documentclass[reprint,superscriptaddress,amsmath,amssymb,aps,prb]{revtex4-1}

\usepackage{color}
\usepackage{graphicx}
\usepackage{dcolumn}
\usepackage{bm}
\usepackage[colorlinks=true
,urlcolor=blue
,anchorcolor=blue
,citecolor=blue
,filecolor=blue
,linkcolor=blue
,menucolor=blue
,pagecolor=blue
,linktocpage=true
,pdfproducer=medialab
]{hyperref}

\begin{document}

%\preprint{APS}

\title{Order in a Spatially Anisotropic Triangular Antiferromagnet}
\author{Sedigh Ghamari}
 \affiliation{Department of Physics \& Astronomy, McMaster University\\1280 Main St. W., Hamilton ON L8S 4M1, Canada.}

\author{Catherine Kallin}
 \affiliation{Department of Physics \& Astronomy, McMaster University\\1280 Main St. W., Hamilton ON L8S 4M1, Canada.}

\author{Sung-Sik Lee}
 \affiliation{Department of Physics \& Astronomy, McMaster University\\1280 Main St. W., Hamilton ON L8S 4M1, Canada.}
 \affiliation{Perimeter Institute for Theoretical Physics\\31 Caroline St. N., Waterloo ON N2L 2Y5, Canada.}

\author{Erik S. S{\o}rensen}
 \affiliation{Department of Physics \& Astronomy, McMaster University\\1280 Main St. W., Hamilton ON L8S 4M1, Canada.}
\date{\today}
%
%
%
%---------------------------------------------ABSTRACT----------------------------------------------
\begin{abstract}
The phase diagram of the spin-1/2 Heisenberg antiferromagnet on an anisotropic triangular lattice of weakly coupled chains, a model relevant to Cs$_2$CuCl$_4$, is investigated using a renormalization group analysis which includes marginal couplings important for connecting to numerical studies of this model. In particular, the relative stability of incommensurate spiral spin density order and collinear antiferromagnetic order is studied. While incommensurate spiral order is found to exist over most of the phase diagram in the presence of a Dzyaloshinskii-Moriya (DM) interaction, at small interchain and extremely weak DM couplings, collinear antiferromagnetic order can survive.  Our results imply that Cs$_2$CuCl$_4$ is well within the part of the phase diagram where spiral order is stable. The implications of the renormalization group analysis for numerical studies, many of which have found spin-liquid like behavior, are discussed.
\end{abstract}
%
%
%
%---------------------------------------------------------------------------------------------------
%
%
%
\maketitle
%__________________________________________________________________________________________________________________________
%
%
%
%
%---------------------------------------------------------Introduction------------------------------------------------------

\section{Introduction}
\label{intro}

Frustrated magnets are of considerable interest because of their potential for exhibiting novel ground states in two or higher dimensions.\cite{*[{See, for example, }][{ and references therein.}] frus} The spin-1/2 Heisenberg antiferromagnet (HAF) on an anisotropic triangular lattice is a particularly well-studied model of frustrated magnetism and is relevant to Cs$_2$CuCl$_4$, a material whose neutron spectrum has been taken as evidence for close proximity to a two-dimensional spin liquid phase.\cite{coldea, coldea_2, YBK} Below 0.6K, Cs$_2$CuCl$_4$ orders into an incommensurate spiral spin density state and is well described by the spin-1/2 HAF on a triangular lattice, with interchain diagonal exchange $J'$ weaker than the intrachain exchange $J$, although weak interlayer and Dzyaloshinskii-Moriya (DM) interactions also play a role in stabilizing the ordered state. 

The spiral phase with ordering wave vector $q_{cl}=\pi+2\sin^{-1}(J'/2J)$ is the classical ground state for the HAF on a triangular lattice for all $J'< 2J$. It is well established that the nearly isotropic ($J'\lesssim J$) spin-1/2 HAF exhibits spiral order that smoothly connects to the three sublattice N{\'e}el order at $J'$=$J$. Several studies have proposed that, as $J'$ is further reduced, quantum fluctuations destroy this order and stabilize a two-dimensional spin liquid phase.\cite{sl,sl1,sl2,YBK,yunoki,shen,MSW} Although this model is in proximity to a one-dimensional spin liquid ($J'$=$0$), it lacks the features typically associated with possible spin liquid order in higher than one dimension. That is, it neither exhibits macroscopic degeneracy of the classical ground state, as occurs for Kagome or pyrochlore lattices, nor substantial ring exchange, as occurs near a metal-insulator transition. Nevertheless, essentially all calculations show that the quantum fluctuations are unusually large and the local moment has been estimated to go to zero for $J'/J$ as large as $0.9$.\cite{MSW,dalidovich,sorella,thomale}

While it has typically been assumed that the loss of spiral order signals the appearance of a quantum disordered state, Starykh and Balents,\cite{sb} using a renormalization group (RG) approach, found a collinear antiferromagnetic (CAF) state stabilized at small $J'/J$.\cite{sb,sb2} They identified quantum fluctuations of order $(J'/J)^4$ that couple second nearest-neighbor (nn) chains antiferromagnetically, leading to CAF order, whereas a ferromagnetic second nn chain interaction would be compatible with spiral order. This result is surprising since, although quantum fluctuations will typically select a specific state from a degenerate manifold of classical ground states, it is unusual for fluctuations to stabilize an ordered state of higher classical energy.\cite{[{Similar physics has been proposed in another system. See }]turner} This order is rather subtle since it is only selected at $\mathcal{O}(J'/J)^4$, so numerical studies taken as evidence for spin-liquid behavior may also be compatible with such order, as well as with spiral order which is strongly renormalized by quantum fluctuations. Exponentially weak spiral order has been predicted within a random phase approximation\cite{tsvelik} which does not include the $(J'/J)^4$ antiferromagnetic fluctuations identified in Ref.~\onlinecite{sb}. Numerical studies on multi-leg ladders find incommensurate spiral correlations with a wavelength which is strongly renormalized from the classical value and smaller than the finite system size for $0.5\le J'/J<1$.\cite{dmrg} This provides evidence for an incommensurate spiral ground state in this range of $J'$, but one would need to study even larger systems for smaller $J'$ since the spiral wavelength, $\lambda=2\pi/(q-\pi)$, is expected to grow rapidly with decreasing $J'/J$. 

Calculations that directly compare the energies of the CAF and spiral states have given conflicting results. Pardini and Singh,\cite{singh} using linked cluster series expansions, found that the spiral phase, with a variational ordering wave vector, appears to have lower energy than the collinear state for $J'/J\geq 0.1$. They also found the ordered moment vanished rapidly with decreasing $J'/J$, as expected for either ordered state. By contrast, a coupled cluster method yielded a lower energy for the CAF state for $J'/J\lesssim 0.6$, although this method was unable to capture the vanishing moment.\cite{bishop1,bishop2} Neither method is strictly variational, making direct comparisons of the energies difficult to interpret, as discussed in Ref.~\onlinecite{singh}. Nevertheless, both methods show that the energy difference between these two states at small $J'/J$ is significantly smaller than the energy scale set by quantum fluctuations, i.e., than the $\mathcal{O}(J'^2/J)$ difference between the classical and quantum energies.

Here, we reexamine the RG approach of Ref.~\onlinecite{sb}, but rather than starting by integrating out every second chain as was done in Ref.~\onlinecite{sb}, we adhere more strictly to the real-space RG method and keep quantum fluctuations of $\mathcal{O}(J'/J)^{2}$. These are generated by marginal (and irrelevant) couplings and are not expected to order the system as $J'\rightarrow 0$. However, these fluctuations dominate the energy and are important for interpreting numerical results as they extend over an unusually large length scale (which diverges as $J'\rightarrow 0$). Surprisingly, these are of the opposite sign to the fluctuations responsible for CAF ordering.   Thus, our RG analysis elucidates the competition between spiral and CAF order. We also present numerical studies that show that ferromagnetic fluctuations dominate at all $J'<J$, and, consistent with the RG analysis, these fluctuations only grow weakly with system size for small systems. 

The remainder of this paper is organized as follows. The model and RG equations is presented in Sec.~\ref{model_beta-funcs} and the general RG flows and resulting ground states for varying initial conditions are discussed in Sec.~\ref{RG}. Section~\ref{initial_cond} discusses the specific initial conditions appropriate for the Heisenberg model and the resulting RG analysis. The crossover from CAF to spiral order with increasing $J'$ is also discussed in Sec.~\ref{initial_cond}. Numerical studies of this system are presented in Sec.~\ref{numerics} and compared to the RG results, while Sec.~\ref{DM} considers the effect of a weak Dzyaloshinskii-Moriya interaction. A summary and discussion of the results are given in Sec.~\ref{conclusions}. An approximate analytic solution to the RG equations, used in the analysis of the numerical studies, is given in Appendix.

%
%
%
%
%-----------------------------------------------------Model and RG Equations------------------------------------------------
\section{Model and RG Equations}
\label{model_beta-funcs}

\begin{figure}[h]
	\includegraphics[width=1.3in]{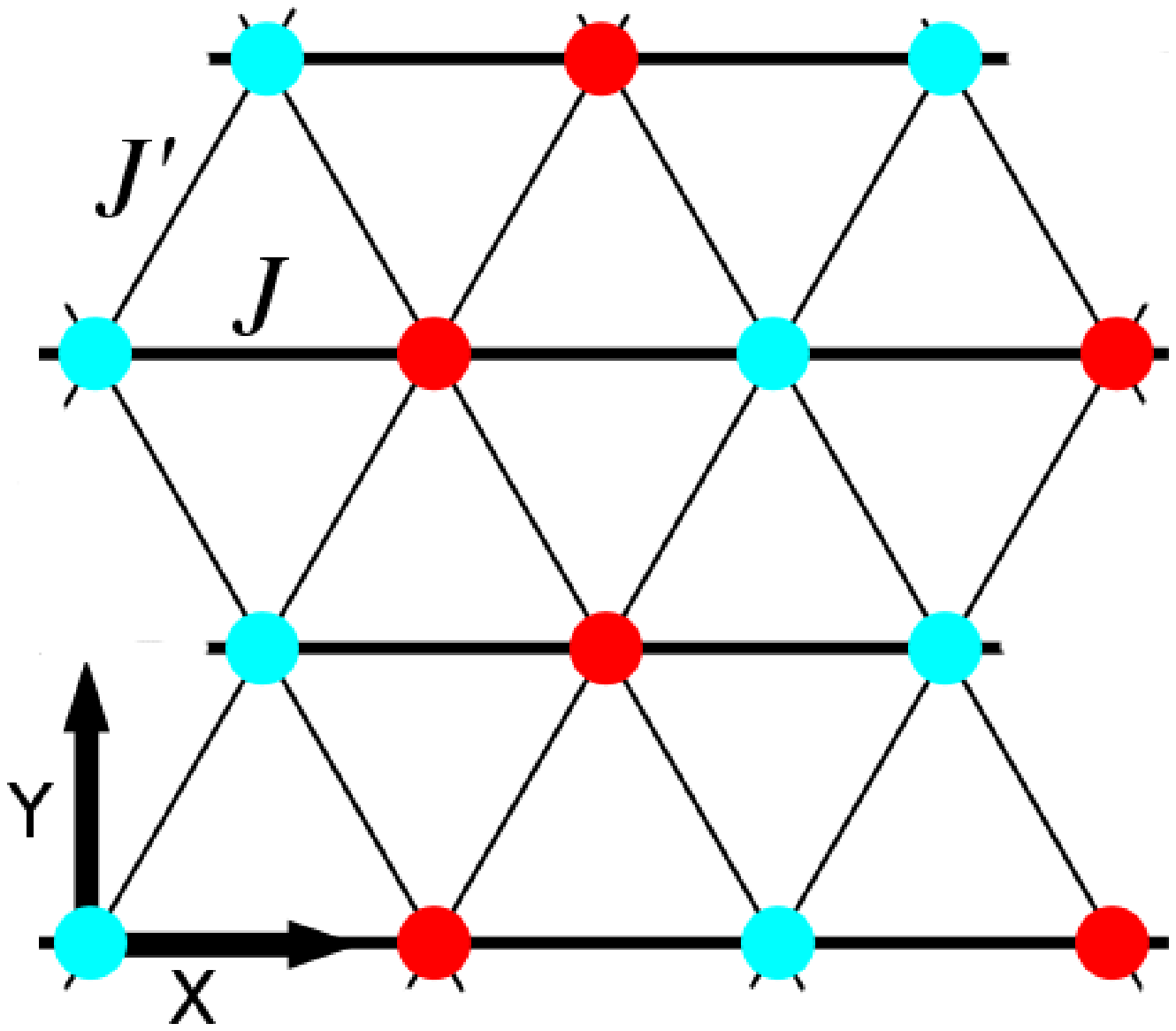}\hspace{0.3in}
	\includegraphics[width=1in , height=1.1in]{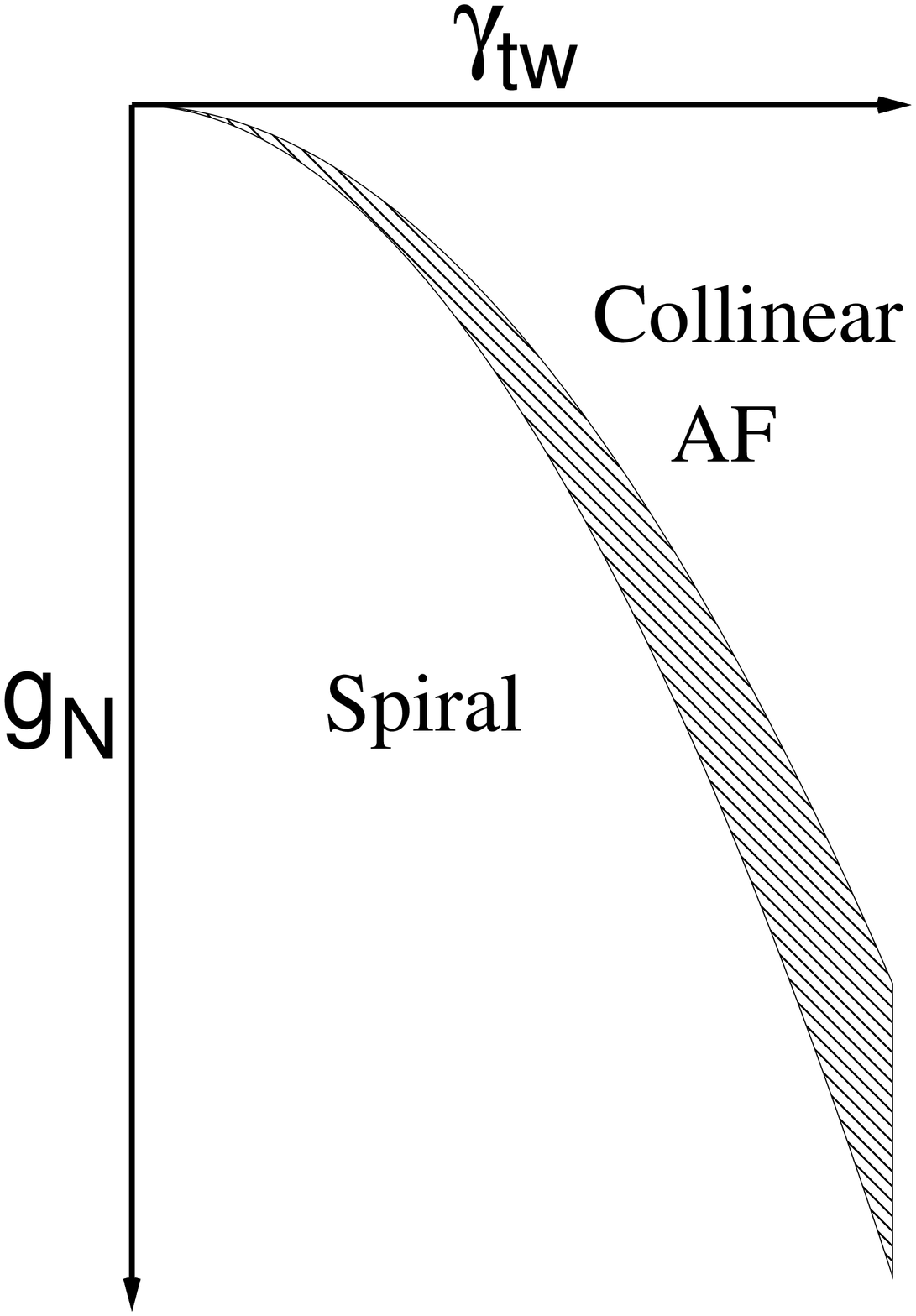}	
	
	(a) \hspace{1.2in} (b)
	\caption{\label{fig_TriLat} (a) Triangular lattice with anisotropic exchanges $J$ and $J'$. Staggered magnetization operators, $\mathbf{N}$, at sites with the same color are accompanied with the same sign consistent with the depicted choice of $y$ axis. (b) Phase diagram projected onto the plane of initial values of the relevant couplings, $g_{N}-\gamma_{tw}$. Dimer or other ordered states may be stable in the shaded region.}
\end{figure}

Here, we derive the RG equations following an approach similar to Ref.~\onlinecite{sb}, but starting from short length scales and including marginal couplings and fluctuations of $\mathcal{O}(J'/J)^{2}$ in order to connect to finite size studies and to study the relative stability of collinear, spiral and dimer ordered states, as summarized in Fig.~\ref{fig_TriLat}. Our starting point is the Hamiltonian:

\begin{eqnarray}
	\mathcal{H} &=& \sum_{n,y} 
	\Big\{ 
		J\, \mathbf{S}_{n,y}\cdot\mathbf{S}_{n+1,y} \, + 
	\label{ham}
	\\
	\nonumber \\
	&& \qquad
		J'\, \mathbf{S}_{n,y}\cdot \big( \mathbf{S}_{n-\tfrac{1}{2},y+1} + \mathbf{S}_{n+\tfrac{1}{2},y+1} \big) \, +
	\nonumber \\
	\nonumber \\
	&& \qquad 
		\mathbf{D}\cdot \Big[ \mathbf{S}_{n,y} \times \big( \mathbf{S}_{n-\tfrac{1}{2},y+1} - \mathbf{S}_{n+\tfrac{1}{2},y+1} \big) \Big]	
	\Big\}
	\nonumber 
\end{eqnarray}
where $n$ indexes sites on the horizontal chains, $y$ is the chain label, and $J$ and $J'$ are, respectively, intra- and interchain couplings as shown in Fig.~\ref{fig_TriLat}\textcolor{blue}{a}. The last term in Eq.~(\ref{ham}) is the interchain DM interaction, where $\mathbf{D} = D{\bf{\hat{z}}}$.\cite{dalidovich,sb2}

We first consider the case $\mathbf{D}$=$0$. The continuum approximation is made for the horizontal chains, while the other direction is kept discrete. The spin operators are written in terms of the \textit{uniform} and \textit{staggered} magnetizations:~\cite{shelton} $\mathbf{S}_{n,y}\rightarrow\mathbf{M}_{y}(x)+(-1)^{n}\mathbf{N}_{y}(x)$, where the subscript $y$ indexes the chains, and we use the sign convention shown in Fig.~\ref{fig_TriLat}\textcolor{blue}{a}.

The Hamiltonian for each chain is the Wess-Zumino-Novikov-Witten (\textit{WZNW}) $\text{SU(2)}$ model which defines the fixed point of decoupled chains, perturbed by intrachain backscattering, with coupling $\gamma_{bs}$:\cite{affleck} 

\begin{eqnarray}  
	\mathcal{H}^{\text{\tiny intra}}&=&\sum_{y}\{\mathcal{H}^{\text{\tiny WZNW}}_{y}
	+\gamma_{bs} \int dx\,\mathbf{J}_{R,y}\cdot\mathbf{J}_{L,y}\}\,\,.
\end{eqnarray}
The interchain part of $\mathcal{H}$ in the continuum limit becomes:

\begin{eqnarray}
	\mathcal{H}^{\text{\tiny inter}}&=&
	\sum_{y}\int dx 
	\{
		\gamma_{M}\mathbf{M}_{y}\cdot\mathbf{M}_{y+1}\,+\,\gamma_{tw}\tfrac{(-1)^{y}}{2}
	\label{hamiltonian}
	\\
	&&
		\big(\mathbf{N}_{y}\cdot\partial_{x}\mathbf{N}_{y+1} -\partial_{x}\mathbf{N}_{y}\cdot\mathbf{N}_{y+1}\big) + \cdots
	\}\, . \nonumber
\end{eqnarray}
where ellipses indicate irrelevant terms with more derivatives.
$\mathbf{M}_{y}$ is the sum of the left- and right-moving currents, $\mathbf{J}_{L,y}$ and $\mathbf{J}_{R,y}$, which have scaling dimension $1$, so $\mathbf{M}_{y}\cdot\mathbf{M}_{y+1}$ and backscattering both have scaling dimension $2$ and are marginal. The scaling dimension of $\mathbf{N}$ is $1/2$ and, therefore, that of the twist term (with coupling $\gamma_{tw}$) is $2$. 

In addition to the above couplings, there are two relevant and two marginal interactions that are not prohibited by any of the symmetries of this model: 
$ g_{N}\mathbf{N}_{y}\cdot\mathbf{N}_{y+2} $ and $g_{\varepsilon}\varepsilon_{y}\varepsilon_{y+2}$ 
with scaling dimension $1$, and $\gamma_{\varepsilon}(-1)^{y} (\varepsilon_{y}\partial_{x}\varepsilon_{y+1}-\partial_{x}\varepsilon_{y}\varepsilon_{y+1})/2$ 
 and $g_{M}\mathbf{M}_{y}\cdot\mathbf{M}_{y+2}$
with dimension $2$. $\varepsilon$ is the staggered dimerization operator and has scaling dimension $1/2$.\cite{bosonization} 
Third and further nn chain couplings can be ignored for small $J'/J$ because of their much smaller initial values.

The theory is regularized by imposing a short distance cutoff for how close to each other the operators that perturb $\mathcal{H}^{\text{\tiny WZNW}}$ can be.  Real-space RG along $x$ is then conveniently performed using operator product expansions (OPE) of the above interaction terms at separate space-time points and integrating over short relative spatial and temporal separations.\cite{senechal, balents2} The OPEs of chiral currents can be directly derived from the OPEs of chiral fermion fields, but the correct OPEs of $\mathbf{N}$ and $\varepsilon$ require bosonization.\cite{shelton} Neglecting terms of $\mathcal{O}(J'/J)^{5}$ and higher, the $\beta$-functions for the relevant and marginal coupling constants are:~\cite{[{The only omitted marginal $\mathcal{O}(J'/J)^4$ term in the presented set of $\beta$-functions is a long-range analogue of $\gamma_{tw}$ that couples $\mathbf{N}$ and $\partial_{x}\mathbf{N}$ on chains three chains apart. The inclusion of this term in the $\beta$-functions only introduces a small quantitative change and has no qualitative effect on the behaviour of the running couplings}] dummy}

\begin{eqnarray}
	 \partial_{l}\gamma_{bs}&=&\gamma^{2}_{bs}-6g^{2}_{N} \, ,
		\qquad \qquad 
  	 \partial_{l}\gamma_{M}=\gamma^{2}_{M}
	 \label{gbs} \\
	 \nonumber\\
	 \partial_{l}\gamma_{tw}&=&-\tfrac{1}{2}\gamma_{bs}\gamma_{tw}+\gamma_{M}\gamma_{tw}
				  -3\gamma_{tw}g_{N}-\tfrac{1}{2}\gamma_{M}\gamma_{\varepsilon}\label{gtw}
	 \\
	 \nonumber\\
	 \partial_{l}g_{N}&=&g_{N}-\tfrac{1}{2}\gamma_{bs}g_{N}
			  +g_M\zeta_N+\tfrac{1}{4}\gamma^{2}_{tw}+g_{M}g_{N}
	 \label{gN} \\
	 \nonumber \\
	 \partial_{l}g_{\varepsilon}&=&g_{\varepsilon}+\tfrac{3}{2}\gamma_{bs}g_{\varepsilon}
				    -\tfrac{3}{2}g_{M}\zeta_{N}+\tfrac{1}{4}\gamma^{2}_{\varepsilon}-\tfrac{3}{2}g_{M}g_{N} 			    
	 \label{gep} \\
	 \nonumber\\
	 \partial_{l}\gamma_{\varepsilon}&=&\tfrac{3}{2}\gamma_{bs}\gamma_{\varepsilon}
					  -\tfrac{3}{2}\gamma_{tw}\gamma_{M}
					  -\tfrac{3}{2}\gamma_{\varepsilon}g_{\varepsilon}\, \label{ge}
	 \\
	 \nonumber\\
	 \partial_{l}g_{M}&=&g^{2}_{M} - \tfrac{1}{4\pi^2}\gamma_{bs}\zeta_{M}
	 \label{gm} \\
	 \nonumber \\
	 \partial_{l}\zeta_{N}&=& -\zeta_{N} - \tfrac{1}{2}\gamma_{bs}\zeta_{N} - \gamma_{tw}^{2} + g_{M} \zeta_{N}
	 \label{zetaN} \\
	 \nonumber \\
	 \partial_{l}\zeta_{M}&=&-2\zeta_M - 8\pi^{2}\gamma^{2}_{M} + \gamma_{bs} \zeta_{M}
	 \label{zetaM}
\end{eqnarray}
All couplings are scaled by $2\pi v = \pi^2 J$, and $e^l=L/L_0$, so that $l(L_0)=0$, where $L$ is the physical length and $L_0$ is the initial length scale, which has been set equal to one in the above equations. 
Equations.~(\ref{gm})-(\ref{zetaM}) are only brought in because these couplings affect the initial values of more relevant couplings. $\zeta_{N}$ is the coupling for the interaction term $\partial_{x}{\bf N}_y\cdot\partial_{x}{\bf N}_{y+2}$, and $\zeta_M$ is that of $(\mathbf{M}_{y}\cdot\mathbf{M}_{y+2})\,(\mathbf{J}_{L,y+1}\cdot\mathbf{J}_{R,y+1})$. Both of these are irrelevant couplings with scaling dimensions 3 and 4. The $\beta$-functions of $\gamma_{M}$ and $g_{M}$ only describe the renormalization of the interchain backscattering part of $\mathbf{M}_{y}\cdot\mathbf{M}_{y^{\prime}}$ (coupling between currents with opposite chiralities) which also are the parts that enter in the other $\beta$-functions. 

The $\beta$-functions of Ref.~\onlinecite{sb} differ from Eqs.~(\ref{gbs})-(\ref{ge}) in that only the first three terms in the $\beta$-functions for $g_{N}$ and $g_{\varepsilon}$ [in Eqs.~(\ref{gN})-(\ref{gep})] were included in Ref.~\onlinecite{sb} and the renormalization of marginal couplings other than $\gamma_{bs}$ were not considered. $\gamma_{tw}$, in particular, is needed to interpret the numerical results and to study the relative stability of the spiral and collinear AF states. As discussed below, the $\gamma_{tw}^2$ term in Eq.~(\ref{gN}) is not expected to order the system, but describes the dominant fluctuations at short and intermediate distances and is important for interpreting numerical results on finite systems. The key RG equations for our study are Eqs.~({\ref{gbs})-(\ref{gN}).

%
%
%
%
%-----------------------------------------------------------RG Flows---------------------------------------------------------
\section{General RG flows} 
\label{RG}

At the lattice scale, only couplings present in the bare Hamiltonian, $\gamma_{tw}=\gamma_{M}/2=J'/\pi^2 J$, $\gamma_{bs}= -0.23$,~\cite{[{}] [{. Note that $\gamma_{bs}=-0.23$ is an estimated value at $4a_0$.}] eggert} are nonzero. Integrating out short wavelength fluctuations up to the initial length scale $L_0$ (larger than but comparable to the lattice spacing) generates all couplings allowed by symmetry, in particular, the two relevant couplings, $g_{N}(0)$ and $g_{\varepsilon}(0)$. Once generated, these couplings tend to grow exponentially in $l$ (or linearly in $L$) while the marginal and irrelevant couplings grow at most logarithmically with $L$.  It follows from Eqs.~(\ref{gN}) and (\ref{gep}) that, in the presence of negative backscattering, $g_N$ grows faster than $g_\epsilon$.  Therefore, for small $J'/J$, the value of $g_N(0)$ largely determines the fate of the system, as it will reach one (or $J$ in unscaled units) at some length scale, $l^*$, while the other couplings will remain much smaller.

If $g_N(0)$ is positive (i.e., antiferromagnetic), it follows from Eq.~(\ref{gN}), that $g_N(l)$ remains positive and flows to +1 at some $l^*$. In this case, second nn chains will order antiferromagnetically. This is the result found in Ref.~\onlinecite{sb}, where every second chain was integrated out to give $g_N(0)=-2g_\epsilon(0)/3=A_0^xJ'^4/\pi^6J^4$, where $A_0^x \approx 0.13$ is a normalization factor. The CAF state is then stabilized for small $J'/J$, by the order-from-disorder mechanism, when one includes the effect of $\gamma_M(l^*)\sim\mathcal{O}(J'/J)$.\cite{sb}

However, numerics on finite systems show that second nn chains are coupled ferromagnetically, not antiferromagnetically as implied by $g_N(0)>0$. Figure~\ref{fig_ED} shows the interchain Neel susceptibility, where a small finite staggered magnetic field is applied to one chain and the response of the second neighbor chain is calculated using exact diagonalization (ED) for three chains of length 12 sites or less, with open boundary conditions.  For a staggered field, $h$, along $\hat z$, the response $\langle S_i^z\rangle/h$, where site $i$ is one of the center sites on the chain, is calculated. This response is the interchain Neel susceptibility, 

\begin{equation}
 \chi_{s}(L) = -i\sum_{n=1}^L (-1)^{n}\int dt\theta(t)\langle [S^{z}_{i,y}(t),S^{z}_{n,y+2}(0)] \rangle \, .
 \label{chi}
\end{equation} 
It is found that $\langle \vec{S}_i\rangle$ aligns ferromagnetically with the second neighbor chain for all $0<J'<1$, as well as for small $J'<0$.  $\chi_s$, which is discussed further in Sec.~\ref{numerics} below, is analytic in the couplings and, for small $J'/J$, is proportional to $g_N(L)$ to leading order. Therefore, it follows from these ED results that $g_N(0)=a(J'/J)^2+b(J'/J)^3+c(J'/J)^4$, where $a,b$, and $c$ are all ferromagnetic and, in fact, ferromagnetic fluctuations dominate in these systems of up to 36 spins. $\chi_s$ calculated for four chains with periodic boundary conditions applied perpendicular to the chains (along $y$) yields exactly the same quadratic and cubic terms and a larger ferromagnetic quartic term. The difference in the quartic term is due to the longer range analog of $\gamma_{tw}$ which couples third neighbor chains and is not included in the above RG equations as it has no qualitative effect on the running couplings.~\cite{dummy}  

\begin{figure}[h]
\begin{center}
\includegraphics[angle=-90 , width=3.3in]{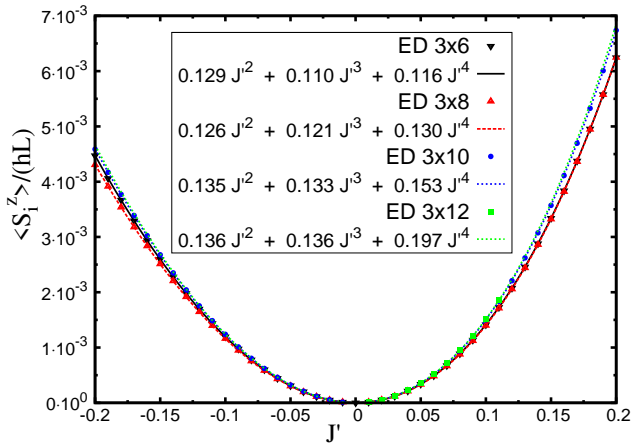}
\caption{\label{fig_ED} The response to a small staggered magnetic field, $h$, applied to one chain is studied using exact diagonalization for three chains of lengths from $L=6$ to $L=12$. The response of a central spin on the second-neighbor chain, $\langle S_i^z\rangle/hL$, is shown along with polynomial fits to the data. The sign of the response is found to be such that $\langle S_i^z\rangle$ always aligns ferromagnetically with the other chain. }
\end{center}
\end{figure}
 
The calculated $\chi_s$ is sensitive to boundary conditions for small systems. Periodic boundary conditions (PBC) along the chains (along $x$) frustrate spiral correlations generated by $\gamma_{tw}$, which we will see in the next section are connected to ferromagnetism. Indeed, for the small systems studied with PBC along $x$ (up to 30 spins, not shown here) we found $\chi_s$ is of order $(J'/J)^4$ and has the opposite sign to the data in Fig.~\ref{fig_ED}. However, the fact that spins $\vec S_{i,y}$ and $\vec S_{i+x,y+2}$ for small $s$ are coupled ferromagnetically at order $(J'/J)^2$ and the higher one is independent of the boundary condition and is seen even in small systems with PBC along $x$. In particular,  $\langle \vec S_{i,y}\cdot \vec S_{i,y+2}\rangle$ is always positive irrespective of the boundary condition.  This implies that $g_N(0)<0$.  In fact, for open boundary conditions, our ED studies find that $(-1)^x\langle \vec S_{i,y}\cdot \vec S_{i+x,y+2}\rangle$ is positive for all $x$, all systems sizes and $J'$ values studied ($J'/J<0.5$).

It follows from the RG equations that for $g_N(0)<0$, $g_N(l^*)$ can be either negative or positive, depending on the precise value of $g_N(0)$. In this case, the flow of $g_N$ is sensitive to the magnitude of the initial value due to the competition between $g_N(0)<0$ and terms in Eq.~(\ref{gN}) that drive $g_N(l)$ antiferromagnetic. This sensitivity to initial conditions can be seen by numerically studying the RG flows for an initial $g_N(0)=\alpha\gamma_{tw}^2(0)$ as the constant $\alpha$ is varied. Here, we use bare values (i.e., values at the lattice scale) for the initial values of the other couplings. 
RG flows are shown in Fig.~\ref{fig_RG} for $-0.30\lesssim\alpha\lesssim-0.26$. For $\alpha\gtrsim -0.28$, one finds that $g_{N}$ flows to 1 (strong antiferromagnetic coupling) and $\gamma_{tw}$ remains small, while for $\alpha\lesssim -0.28$, $g_{N}$ flows to -1 (strong ferromagnetic coupling) and  $\gamma_{tw}$ increases marginally. As discussed below, $g_{N} \to -1$ signals spiral order, whereas $g_{N} \to 1$ signals CAF order as studied by Starykh and Balents.\cite{sb}  However, as seen in the inset of Fig.~\ref{fig_RG} for $g_{N} \to 1$, even in the case of CAF order, the initial RG flows can display increasing ferromagnetic coupling between second nn chains. In Sec.~\ref{initial_cond}, it will be shown that this is the case for the Heisenberg model. In a narrow range for $\alpha\approx -0.28$ (which widens as $J'$ is increased), the growth of $g_{N}$ is hindered and $\gamma_{tw}$, $\gamma_M$ or $g_{\varepsilon}$ reach unity first. This crossover value, $g_N^{\text{\tiny crit}}$, depends on $\gamma_{bs}$ and $J'$. Depending on initial conditions, columnar or staggered dimerizations or more complicated incommensurate states can be stabilized in this crossover region, as denoted by the shaded area in Fig.~\ref{fig_TriLat}\textcolor{blue}{b}.

\begin{figure}[b]
	\begin{center}
		\includegraphics[angle=-90,width=3.4in]{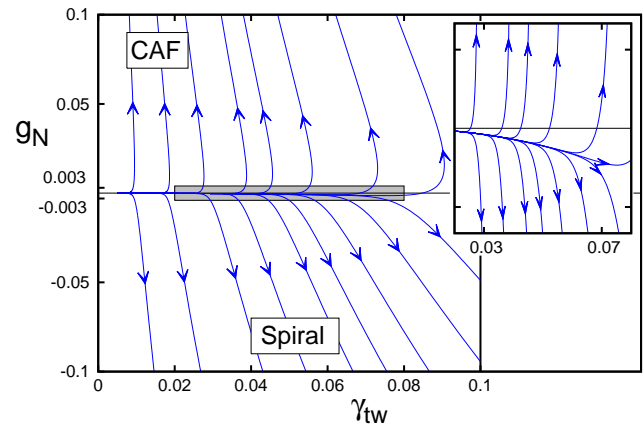}
		\caption{\label{fig_RG} Flows of $g_{N}(l)$ and $\gamma_{tw}(l)$ for several initial values with ferromagnetic $g_N(0)=\alpha \gamma_{tw}^2(0)$, $g_{\varepsilon}=\gamma_{\varepsilon}=0$, and $J'=0.05$. $g_N$ and $\gamma_{tw}$ are in scaled units (divided by $\pi^2 J$), and $\alpha$ is varied from -0.26 (CAF flow closest to vertical axis) to -0.30 (spiral flow closest to vertical axis). The inset zooms in on the shaded rectangle. Flows emanating from initial values very close to $g_{N}^{\rm crit}$ have flow lines substantially away from the $\gamma_{tw}=0$ axes and grow considerably slower than those close to the vertical axes.  Note the extreme sensitivity to initial values close to the critical value $g_{N}^{\rm crit}$ as shown in the inset.}
	\end{center}
\end{figure}

In the ferromagnetic regime where $g_{N}$ grows exponentially to negative values ($g_N<g_{N}^{\rm crit}$), sites along the $y$ direction align, consistent with zero $y$ component of the ordering wave vector. In this regime, $\gamma_{tw}$, which stabilizes spiral order, grows marginally. In the classical limit, the ordering wave vector $\mathbf{q}\equiv(\pi+\epsilon)\mathbf{\hat{x}}$ is $(\pi+J'/J)\mathbf{\hat{x}}$ for $J'<<J$. The effect of quantum fluctuations on $\mathbf{q}$, or $\epsilon$, can be determined from the RG analysis. The most robust spiral state occurs if ferromagnetism is selected at quadratic order in $J'$. The length scale, $l^{*}$, at which $g_{N}$ becomes comparable to $J$ in magnitude is inversely proportional to the initial value of $g_{N}$ which, in turn, is $\propto J'^{2}$ in this case. Since $\gamma_{bs}$ changes logarithmically and remains negative as $g_N$ grows, at this length scale, we expect almost the same intrachain Hamiltonian but with a somewhat renormalized $\gamma_{tw}$. This Hamiltonian describes interactions between blocks of N{\'e}el-ordered spins of length $l^{*}$ and can be treated classically. Thus, the ground state has a spiral order with $\epsilon \propto (J'/J)^{3}$. Since $1/\epsilon\gg l^{*}$ for small $J'$, the pitch of the spiral is much longer than the Neel blocks and this treatment is justified. This is a significant renormalization of the ordering wave vector toward the 1D limit of $\pi$, where spins on neighboring antiferromagnetic chains are oriented at $\pi/2$ with respect to each other and requires that ferromagnetism is selected at quadratic order. 

Even more fragile spiral order is stabilized if $\Delta g=g_{N}(0)-g_{N}^{\rm crit}\approx 0$ and ferromagnetism is selected at cubic or quartic order in $J'$. Then $\epsilon \propto (J'/J)^{n}$, where $n=4$ (or 5) for cubic (or quartic) selection. The exponentially weak spiral order, $\epsilon \sim e^{-a(J/J')^2}$ found within a random phase approximation,\cite{tsvelik} also follows from the RG equations if one assumes that neither $g_N$ nor $g_{\epsilon}$ can grow. To leading order in $J'/J$, Eq.~(\ref{gtw}) describes this exponentially weak spiral order, where $a=2.6$ if the value $\gamma_{bs}\approx -0.38$ at $a_0$ is used.  However, since there is no symmetry that prevents $g_N$ (or $g_\epsilon$) from growing at all orders, one does not expect an exponentially weak spiral order to be stable for very small $J'/J$.  

As $J'$ increases, even for initial conditions that favor antiferromagnetism, $\Delta g=g_{N}(0)-g_{N}^{\rm crit}<0$, the RG flows move toward increasing $\gamma_{tw}$, i.e., toward the spiral state. This is because $\gamma_{tw}$ is boosted by both backscattering and $\gamma_M$. The latter enhances the spiral incommensuration, $\epsilon$, so it is larger than $e^{-a(J/J')^2}$. For the Heisenberg model, where $\gamma_{bs}(0)$ is sufficiently negative to suppress the growth of $g_\epsilon$ relative to $g_N$, the competition is predominantly between the CAF and the spiral states. If a sufficiently strong frustrating second nn interaction along the chains is added to the Hamiltonian, the competition is then between the CAF and dimerized phases as discussed in Ref.~\onlinecite{sb}.

%
%
%
%
%----------------------------------------------------Initial Conditions---------------------------------------------

\section{Initial Conditions and RG Results} 
\label{initial_cond}

It follows from the above discussion of the RG flows that it is important to determine $g_N(0)$ accurately. Numerics on small systems clearly show that $g_N(0)$ is ferromagnetic, whereas Eq.~(\ref{gN}) would naively predict that $g_{N}(0) \sim + \gamma_{tw}^2$. In fact, only antiferromagnetism appears to enter the RG equation for $g_N$ to all orders in $J'/J$. This would imply not only that spins on second neighbor chains are always antiferromagnetically ordered, but that the dominant fluctuations at all length scales are antiferromagnetic.  Actually, one sees how second neighbor chains can be coupled ferromagnetically by considering the effect of a twist in one chain on spins in neighboring chains. Treating $\gamma_{tw} \mathbf{N}_{y\pm 1} \cdot \partial_{x} \mathbf{N}_{y}$ as perturbations to the decoupled 1D chains, one finds in second order that fluctuations in the $y$-th chain generate a local ferromagnetic coupling, $\mathcal{H}^{\text{\tiny eff}} \sim - \gamma_{tw}^{2} \langle | \partial_{x} \mathbf{N}_{y} |^{2} \rangle \mathbf{N}_{y-1} \cdot \mathbf{N}_{y+1}$. Although the static twist $\langle \partial_{x} \mathbf{N}_{y} \rangle$ is small, the fluctuation can be $\mathcal{O}(1)$, mediating an interaction of order $J'^{2}$. 

The peculiar behavior where the sign of the quantum correction
for $g_N$ depends strongly on scale can also be seen in the continuum theory,
once the theory is regularized.
The OPE between two twist terms generates a term in the effective action: 

\begin{equation}
	-\gamma_{tw}^2 \int dz dz' 
	\left[ 
	\partial_{x} \partial_{x'} 
	G(z-z')
	\right]
	{\bf N}_{y-1}(z)
	\cdot
	{\bf N}_{y+1}(z').
	\label{Seff}
\end{equation}
Here $z=i(vt-x)$, $G(z-z') = \langle \mathbf{N}_{y}(z)\cdot\mathbf{N}_{y}(z') \rangle \sim 1/{\sqrt{ (z-z')^2 + a_c^{2} }}$ and
$a_c$ is larger than but comparable to the lattice spacing. The integrand in Eq.~(\ref{Seff}) is positive for $0 < |z-z'| < a_c$, which generates the ferromagnetic initial condition for $g_N$, but becomes negative for $|z-z'| > a_{c}$, favoring antiferromagnetic correlation at longer lengths. 

The presence of a ferromagnetic coupling at short distances is very general and independent of the specific regularization scheme: it simply follows from the fact that $\left[ \partial_{x} \partial_{x'} G(z-z')\right]$ in Eq. (\ref{Seff}) is a total derivative which integrates to zero and is negative at large distances. The presence of this total derivative implies that this term does not contribute at $q=0$ in a gradient expansion and, consequently, is not expected to order the system.\cite{[{We thank Leon Balents and Oleg Starykh for pointing this out}]leon} For example, if one fully integrates out every second chain to generate effective couplings between second nn chains, as was done in Ref.~\onlinecite{sb}, this term vanishes. This implies that the initial value of $g_N$ at quadratic order must be such as to cancel the effect of $\gamma_{tw}^2$ in the $\beta$-function in the $q \to 0$ limit. This places constraints on the initial conditions that are discussed below.  In general, however, $g_N(0)=aJ'^2+bJ'^3+cJ'^4$, where $a,b,$ and $c$ are all negative (\textit{i.e.}, ferromagnetic).  The cubic and quartic contributions to the initial condition are related to the quadratic and cubic contributions in the beta function for $\gamma_{tw}$.  

We note here that the $\gamma_{tw}$ term in the $\beta$-function for $g_N$ is not the only term that contributes ferromagnetism at short length scales. For example, the longer range analog of $\gamma_{tw}$ and irrelevant couplings such as $\zeta_N$ (or even higher derivative terms) contribute to the $\beta$-function for $g_N$ in a very similar way to the $\gamma_{tw}^2$ term, but with smaller coefficients. While the inclusion of such irrelevant terms can have a quantitative effect at short lengths, the qualitative behavior of strong ferromagnetism, which persists to fairly long-length scales, is captured by including the effects of the marginal coupling, $\gamma_{tw}$, only and is unchanged by the inclusion of more irrelevant couplings.  

That $g_N$ should not grow exponentially due to the $\gamma_{tw}^2$ term is not only suggested by Eq.~(\ref{Seff}), but can be seen by expanding the interchain susceptibility, $\chi_s$, in powers of $J'$, treating the interchain interaction as a perturbation. One finds, using the simple power counting suggested by scaling, that $\chi_s(L)/L$ grows at most logarithmically in $L$ at quadratic and cubic orders, but there is a contribution at quartic order that grows linearly in $L$. It follows from this perturbative expansion that the $\gamma_{tw}^2$ term causes no exponential growth in $g_N(l)$. The initial condition that corresponds to ``tuning" (no exponential growth) is given in Eq.~(\ref{gncrit}) in Appendix, where the RG equations are solved analytically to cubic order. For $\gamma_{bs}=0$ and the initial condition, $g_N(0)=-J'^2/4$, it follows from Eqs.~(\ref{gtw}) and (\ref{gN}) that $g_N(l)$ remains constant at quadratic order. However, for nonzero $\gamma_{bs}$, $g_N(l)$ flows even at quadratic order and a larger ferromagnetic initial condition is needed to ensure that $g_N$ does not grow exponentially in $l$ at quadratic order. 

While there is a ferromagnetic quartic contribution to $g_N(0)$ from the $\gamma_{tw}^2$ term, there is also an antiferromagnetic contribution from the $g_{M}\zeta_N$ term in the $\beta$-function. This term, generated within the operator product expansion technique, is analogous to the antiferromagnetic initial condition identified by Starykh and Balents.\cite{sb} $g_{M}\zeta_N$ is zero at the lattice scale since these couplings do not appear in the microscopic Hamiltonian, but is nonzero in subsequent RG steps and provides an initial condition equal to $8\gamma_{bs}J'^4/\pi^9J^4=0.06J'^4/\pi^6J^4$ at $\ell=0$ if one sets $\gamma_M(0)=2\gamma_{tw}(0)=2J'/\pi^2 J$. This differs from the Starykh and Balents initial condition $g_N(0)=A_0^xJ'^4/\pi^6J^4$ by about a factor of two, but both calculations rely on the continuum approximation which is only approximate near the lattice scale. Also, the two results are almost identical if one uses the value of $\gamma_{bs}$ extrapolated to the lattice scale. An analysis of both calculations shows that they rely on the same physics, which involves generation of a gradient coupling between second-neighbor chains ($\zeta_N$) and requires chiral symmetry breaking (i.e., nonzero $\gamma_{bs}$ in the continuum RG language). 

The long-distance RG results of Ref.~\onlinecite{sb} are recovered within our analysis for sufficiently small $J'$ if the ferromagnetic initial condition associated with $\gamma_{tw}^2$ is equal to $g_N^{\rm{crit}}$ to $\mathcal{O}(J'/J)^{4}$. In this case, the quartic term, $g_{M}\zeta_N$, provides the essential contribution to the initial condition of $g_N$ that drives the ordering.  For this initial condition, the RG flow for small $J'$ corresponds to one of the flows shown in Fig.~\ref{fig_ED} where $g_N$ is initially ferromagnetic, but passes through zero at an intermediate length and then becomes large and antiferromagnetic. For small $J'$, the intermediate length where $g_N$ crosses over from ferromagnetic to antiferromagnetic, is proportional to $(J/J')^2$, neglecting logarithmic corrections. This follows from the fact that the quadratic contribution to $g_N(0)$ grows at most logarithmically in $L$, while the quartic antiferromagnetic contribution grows linearly in $L$. Consequently, the second nn chains in finite systems with chains of length $L$ will be ferromagnetically correlated for $L<L_{\text{FM}}=A(J/J')^2$. The coefficient $A$ is estimated to be 20-40 lattice spacings, depending on which value is used for the antiferromagnetic contribution to $g_N(0)$.

A direct transition from CAF to spiral order is expected to be discontinuous, while the RG flows change continuously. Nevertheless, one can extract some information about this transition from the RG analysis. As $J'$ increases, the RG flows move toward larger $\gamma_{tw}$ at $l^*$, which shows a tendency of moving toward spiral order.  From the numerical solution of the $\beta$-functions of Eqs.~(\ref{gbs})-(\ref{ge}), it follows that $J'_c\sim 0.3$, where $J'_c$ is defined as the value at which $\gamma_{tw}(l^{*})$ first reaches $J$ while $|g_N(l^{*})|\le J$. For $\gamma_{bs}(0)=-0.23$, $g_\epsilon(l^{*})$ remains small compared to $J$ at this $J'$, so the RG predicts that there is no intermediate dimer phase between the CAF and spiral ordered phases. This estimate for $J'_c$ results from using the value $A_0^xJ'^4/\pi^6J^4$ for the AF contribution to the initial condition\cite{sb} and setting the ferromagnetic initial condition or tuning as discussed at the end of Appendix.  

Terms higher order than $(J'/J)^4$ and irrelevant couplings which are ignored, as well as exactly where one terminates the RG flows, can all affect $J'_c$ so $J'_c\sim 0.3$ is only a crude estimate. The main reason for the somewhat small value of $J'_c$ is that the CAF state is only selected at quartic order, while the competing marginal coupling, $\gamma_{tw}$ is linear in $J'$ and boosted both by backscattering and by $\gamma_M$, which is also linear in $J'$.  We note that the estimate of $J'_c$ is compatible with the observation of spiral correlations at $J'=0.5$ in numerical studies of multi-leg ladder systems.\cite{dmrg} 

%
%
%
%
%--------------------------------------------------------Numerical Results---------------------------------------------------
\section{Numerical Results}
\label{numerics} 

The ED results shown in Fig.~\ref{fig_ED} show only ferromagnetism at all $J'$, but the system sizes are not large enough to determine whether the strength of the ferromagnetism is changing with increasing system size for small $J'$. To address this, we turn to a more detailed analysis of the numerical studies of the interchain susceptibility, $\chi_s$, using density matrix renormalization (DMRG) to study larger systems. Specifically, we address the question of what initial conditions are compatible with the study of finite systems or, alternatively, are finite size studies compatible with the RG analysis.  Using finite-size scaling, the susceptibility $\chi_s(L, \{g_i(0)\},L_0)$ of the system with size $L=L_0 e^l$ with couplings $\{g_i(0)\}$ defined at scale $L_0$ $(l=0)$ can be related to the susceptibility of the system with size $L_0$ as 

\begin{equation}
\chi_s(L, \{g_i(0)\},L_0) = {L\over L_0}[1-\gamma_{bs}(L_0)l)]^{1/2} \chi_s(L_0,\{g_i(l)\},L_0) \, , \nonumber
\end{equation} 
where the linear-$L$ dependence in the prefactor is due to $\chi_s$ having scaling dimension $1$ in the absence of the backscattering, and the second factor is the contribution of the running backscattering that modifies the scaling dimension of $\chi_s$. For small $g_N(0)$ (small $J'$), $\chi_s(L_0,\{g_i(l)\},L_0)$ is analytic in the couplings and proportional to $g_N(l)$ to leading order, and we have $\chi_{s}(L)\propto (L/L_0)[1-\gamma_{bs}(L_0)l)]^{1/2}g_N(l)$. There is no further quantum correction to this relation to leading order in $J'/J$, since $g_N(l)$ is fully renormalized and includes all quantum fluctuations.

Since three chains captures all contributions to $g_{N}$ and $\chi_s$ to order $(J'/J)^{3}$ as well as the key $(J'/J)^4$ fluctuations that drive CAF order, we confine our studies to three-chain systems. The interchain susceptibility is studied using ED and DMRG for the HAF on three chains with open boundary conditions for 24 to 84 spins ($L$=8 to 28); ED was used for up to 36 spins. Up to $m=2600$, states were kept in the DMRG, while typically keeping the precision $\epsilon \le10^{-8}$. The results are shown in Fig.~\ref{fig_DMRG}. For all values of $0<J'<J$, $\chi_s$ was found to be ferromagnetic. As noted earlier, ED studies also found that the correlation between second nn chains has quadratic, cubic, quartic (and higher order) in $J'$ contributions which are all ferromagnetic for $L\le 12$ (as expected from analyzing the $\gamma_{tw}^2$ term), and the additional quartic contribution from 4 chains is also ferromagnetic. There is no sign in the data of a turnover to negative values of $\chi_s$ expected for AF correlations. As $J'/J$ increases, the length scale at which antiferromagnetism should be observable if the CAF state is stable becomes shorter.  For example, at $J'/J=0.5$, the RG analysis predicts that one should see a turnover in $\chi_s$ for $L\gtrsim 20$, although we don't expect RG to be quantitatively accurate at such large $J'/J$. As $J'/J$ increases, higher-order fluctuations that are not included in the RG and that are not fully captured by studying three chains, can either further stabilize or destabilize CAF order.  

\begin{figure}[h]
	\includegraphics[angle=-90,width=3.3in]{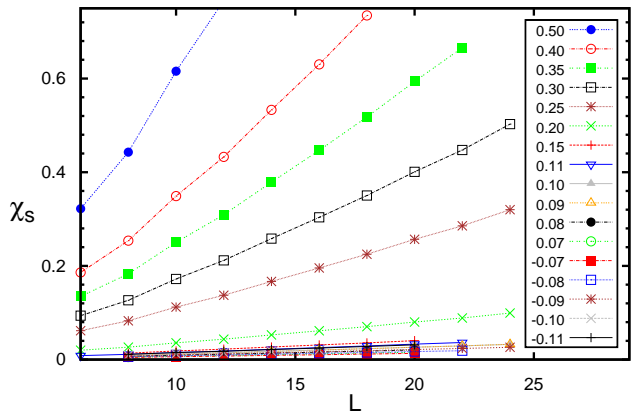}	
	\caption{\label{fig_DMRG} The staggered Neel susceptibility, $\chi_{s}$, for three chains as a function of chain length, $L$, at different $J'$ values, calculated by ED and DMRG. The sign of $\chi_{s}$ is such that second nn chains are ferromagnetically coupled for all $L$ and $J'$ studied.}
\end{figure}

For $L\ge 10$ and $|J'|\leq 0.1J$, the change in $\chi_s$ as a function of $L$ is fit to the predicted RG flow for $g_{N}$, $\gamma_{tw}$, and $\gamma_{bs}$ as functions of initial conditions, treated as fitting parameters. $\gamma_{M}$ is incorporated approximately to capture the cubic behavior of $g_N(l)$ as described in the Appendix. The restriction to small $J'$ is to ensure that $\chi_s$ is proportional to $g_N$ and higher order terms can be neglected. Figure~\ref{fig_Fit} shows $\chi_s$ scaled by $(L/L_0)[1-\gamma_{bs}(L_0)l)]^{1/2}$, with the lines being the RG fits to the data. One can see the different slopes of curves for positive and negative values of $J'$, which shows the effect of the cubic term. It is also clear from these curves that $g_N$ is growing only weakly with $L$. 

The initial conditions at $L=10$ extracted from the numerical data are $\gamma_{tw}=0.416J'+ 0.121{J'}^{2}$, $g_{N}=-0.0450{J'}^{2}-0.0425J'^{3}$, $\gamma_{bs}=-0.072$, and $\gamma_{M}=0.24J'$. While the extracted value of $\gamma_{bs}(L_{0})$ is substantially closer to zero than the value of $-0.19$ for periodic boundary conditions, this is consistent with an independent calculation of $\chi_{s}$ for a single chain. For short chains, open boundary conditions are known to favor dimerization relative to periodic boundary conditions.\cite{affleck3} The fact that the ratio of $\gamma_{tw}(0)/\gamma_{M}(0)$ extracted from the numerical data is noticeably larger than the ratio at the lattice scale can be attributed to the effect of irrelevant couplings that contribute ferromagnetism in an analogous way to the $\gamma_{tw}^2$ term. 

Although the extracted initial conditions are such that $g_N$ flows to a large ferromagnetic value, which would seem to indicate spiral order, the deviation from tuning at both quadratic and cubic orders is sufficiently small to be compatible with finite size effects. For the values of $\gamma_{bs}(0)$, $\gamma_{tw}(0)$, and $\gamma_{M}(0)$ extracted from the data, tuning corresponds to $g_{N}(0)=-0.0447 J'^2-0.0479 J'^3 +\mathcal{O}(J'/J)^4$. Therefore the quadratic (cubic) term deviates by $1\%$ ($12\%$) from the value predicted from the RG equations for an order selected at $\mathcal{O}(J'/J)^4$. Given the uncertainty introduced by finite-size effects for $10<L<30$ and the approximations made in the RG, the numerics for small $J'$ are consistent with no order being selected to cubic order, while the lack of any tendency toward antiferromagnetism with increasing $L$ for $J'/J\ge 0.5$ suggests the CAF state is not stable for these larger values of $J'$. The numerics also confirm the RG prediction that, not only is there ferromagnetism at the lattice scale for all values of $J'/J$, but this ferromagnetism continues to grow to fairly long length scales. Ideally, one would like to study the quartic term for small $J'$ as a function of system size since the RG analysis predicts that CAF order is selected by an initial condition for $g_N$ that deviates from tuning by about 20\% at quartic order.  However, this would require studying larger system sizes while keeping more states in the DMRG calculations and is beyond the scope of this study.

\begin{figure}[h]
	\includegraphics[angle=-90,width=3.3in]{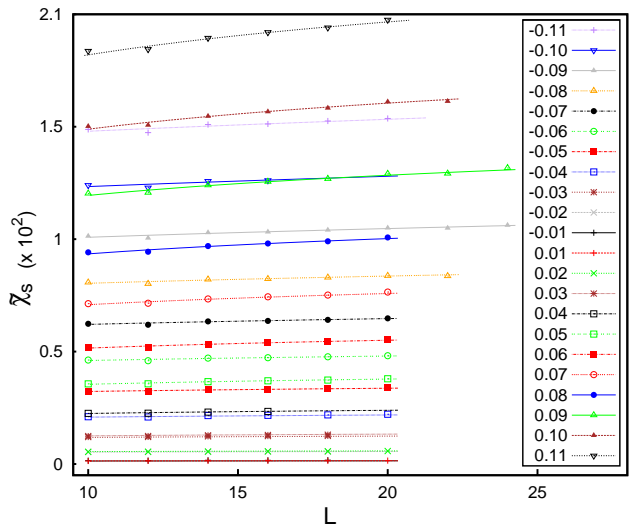}	
	\caption{\label{fig_Fit} The scaled susceptibility, $\tilde{\chi}_{s}=\chi_s/[L(1-\gamma_{bs}(0)l)^{1/2}]$, which is proportional to the relevant coupling, $g_N$, for small $J'$, is shown for different values of $J'/J$.  Points represent data from ED (for $L\le 12$) and DMRG ($L> 12$) and lines are fits to the RG equations, where the initial conditions are used as fitting parameters.} 	
	
\end{figure}

%
%
%
%
%----------------------------------------------------------------DM------------------------------------------------------------
\section{Dzyaloshinskii-Moriya interaction}
\label{DM}

It is also interesting to consider the effect of a nonzero DM interaction in this RG picture, since it is known that the interchain DM interaction further stabilizes spiral order in this model.~\cite{dalidovich,singh2} In the continuum limit, the interchain DM interaction can be written as $\epsilon^{ab}N^{a}_{y}N^{b}_{y+1}$ with the initial condition $g_{\text{D}}(0) = 2\|\mathbf{D}\|=2D/\pi^2 J$. Note that this term breaks the $\text{SU(2)}$ symmetry of the original model and will order the spins in a plane perpendicular to the direction of ${\bf D}$. Just like the $g_N \mathbf{N}_{y}\cdot\mathbf{N}_{y+2}$ term, the DM interaction has scaling dimension $1$ and is relevant. It flows according to the following $\beta$-function:

\begin{equation}
	\partial_{l} g_{\text{D}} \,=\, g_{\text{D}} - \frac{1}{2} \gamma_{bs}g_{\text{D}} - 4 g_{N} g_{\text{D}} + \frac{1}{2} \gamma_{M}g_{\text{D}}\,.
\end{equation} 
The DM interaction also introduces a $-2g_{\text{D}}^2$ term (for components perpendicular to ${\bf D}$) in the $\beta$-function of $g_N$, Eq.~(\ref{gN}).  In addition to getting some boost from the marginal backscattering and $\gamma_M$, the DM interaction both is enhanced by and promotes ferromagnetic $g_N$. Consequently, it suppresses CAF order relative to spiral order. In any region of the phase diagram where spiral order is stable for $D=0$, the DM interaction will further stabilize this phase relative to CAF order.

Since both $g_{\text{D}}$ and $g_N$ have the same exponential growth for small $J'$, a DM interaction with $g_D(0)$ greater than the untuned (antiferromagnetic) part of $g_N(0)$ will overcome the CAF order as $g_{\text{D}}$ will grow to be of $\mathcal{O}(1)$ first. The DM interaction favors neighboring Neel chains to be aligned perpendicular to each other, but in the presence of $\gamma_{tw}$ (which grows marginally), a spiral state will be stabilized. This suggests the phase diagram shown in Fig.~\ref{fig_DM} with a transition between the CAF and spiral states at $D \propto J'^4/J^3$ for small $J'/J$. This transition is expected to be first order, so the RG crossover between these two states is only approximate. However, because CAF order is only weakly favored, at order $(J'/J)^4$, the maximum $D$ for which CAF order survives will also be very small and is estimated from the RG analysis to be less than $10^{-4}J$ if $J'_c\lesssim 0.3J$. Even if the interchain coupling for Cs$_2$CuCl$_4$ is such as to stabilize CAF order in the absence of any DM interaction, the estimated DM interaction for Cs$_2$CuCl$_4$, $D \approx 0.05J$, is much greater than what is required to stabilize spiral order. 

One can also consider the effect of an intrachain DM interaction with the Hamiltonian $\sum_{n,y}\mathbf{D}'\cdot (\mathbf{S}_{n,y} \times \mathbf{S}_{n+1,y})$. This interaction can promote several different phases even in a single spin-1/2 chain for interchain DM coupling, $D'$, comparable to $J$.~\cite{garate} For $D'\ll J$, it has little effect on the phase diagram shown in Fig~\ref{fig_DM}.  One can show that the primary effect of $D'$ is to renormalize the interchain DM coupling, ${\mathbf D}\rightarrow \mathbf{D} - \mathbf{D}'J'/2J$.~\cite{sg}   For $D=0$, this implies that $D'\gtrsim \mathcal{O}(J'^3/J^2)$ will stabilize spiral order at small $J'/J$.  For nonzero $D$ and $|{\mathbf D}'|\sim|{\mathbf D}|$ or smaller, the intrachain DM coupling can slightly tilt the ordering plane and/or slightly shift the phase boundary between CAF and spiral, depending on the relative directions of $\mathbf{D}$ and $\mathbf{ D}'$.  This can be understood from the fact that, for small $J'/J$, the CAF and spiral states exhibit essentially the same AF order along a chain and only differ significantly in how the chains order with respect to each other.

\begin{figure}[h]
	\includegraphics[angle=-90,width=3.2in]{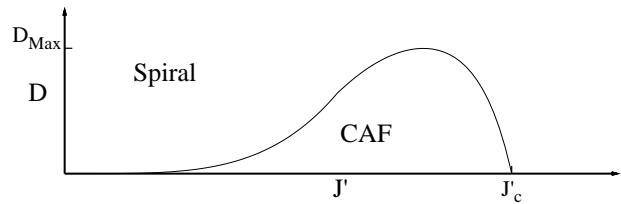}	
	\caption{\label{fig_DM} Suggested phase diagram where the critical interchain Dzyaloshinskii-Moriya coupling, $D$, vanishes as $J'^4/J^3$ at small interchain coupling $J'$. The ordering wave vector of the spiral state varies continuously and is ${\bf q}=(\pi,0)$ for $J'=0$ and nonzero $D$. The region of stability of the CAF state is very small since
	$D_{\rm Max}$ is estimated to be less than $10^{-4}J.$}
\end{figure}

%
%
%
%
%
%-----------------------------------------------------------Summary---------------------------------------------------------
\section{Conclusions and Discussion}
\label{conclusions}

In summary, the full RG equations for the HAF on an anisotropic triangular lattice describe the intense competition between CAF and incommensurate spiral orders. A tiny perturbation of $\mathcal{O}(J'/J)^4$ can tilt the balance between these two states at small $J'/J$, while a larger perturbation is needed to stabilize dimer order. As predicted by Starykh and Balents,\cite{sb} the CAF state is stable at sufficiently small $J'/J$, but either an interchain DM interaction or a ferromagnetic second neighbor interchain interaction of $\mathcal{O}(J'/J)^4$ would stabilize an incommensurate spiral state. By contrast, an $\mathcal{O}(J)$ intrachain interaction that pushes backscattering, $\gamma_{bs}$,  toward zero or an interchain dimer coupling, which is larger than $\mathcal{O}(J'/J)^4$ is needed to stabilize dimer order.  The fact that such an extremely weak interchain DM coupling is sufficient to stabilize spiral order at all $J'/J$, implies that Cs$_2$CuCl$_4$, with an estimated $D \approx 0.05J$, is well within the region of stability for spiral order and is not close to a CAF instability.

For zero or extremely weak DM coupling, the RG analysis suggests a direct transition from CAF to spiral order as $J'$ increases, without an intermediate dimer phase.  This transition is estimated to occur at $J'_c\lesssim 0.3J$ and is compatible with recent DMRG studies which find evidence in favor of spiral order for $J'\ge 0.5J$.\cite {dmrg}  On the other hand, this estimate would seem to be in contradiction to other studies that have concluded that spiral order is lost at significantly larger values of $J'/J$, such as at $J'/J\sim0.9$.\cite{sl1,sorella,thomale}  However, numerics which are interpreted as loss of spiral order could be compatible with the weak spiral order expected in the presence of quantum fluctuations.  Furthermore, since the wavelength of the spiral grows rapidly with decreasing $J'/J$ and its energy lies so close to that of the CAF, as well as to many disordered states, it is important to use open boundary conditions in finite size studies.
 
The RG analysis connects directly to numerical studies on finite systems where, for small $J'/J$, second neighbor chains couple ferromagnetically, the opposite to what one would expect for CAF order.  For the infinite system, the static interchain spin-spin correlation function, $C_2(x)=\langle (-1)^x\vec{S}_{i,y}\cdot \vec{S}_{i+x,y+2}\rangle$, is ferromagnetic and of order $(J'/J)^2$ at $x=0$ even in the CAF state.  The RG analysis suggests that these ferromagnetic correlations persist to quite large $x$ since the ferromagnetic coupling between chains grows with increasing system size up to chain lengths that are smaller than, but comparable to  $L_{\text{FM}}=A (J/J')^2$.  The ED and DMRG results are consistent with this and suggest that the antiferromagnetism is hidden by larger ferromagnetic fluctuations except at very long length scales.  One might be able to see  these ferromagnetic correlations in the CAF state using cluster or series expansion calculations for the infinite system.  

While our numerical studies are consistent with the RG analysis, they do not provide any direct evidence of CAF order.  Given the long-length scales involved for $J'<J'_c$, it would be extremely challenging to identify this order from finite-system-size studies.  For small $J'/J$, a large static magnetic susceptibility has been observed at or very close to $q_x=\pi$,\cite{sl1,thomale} but this feature is common to both CAF and spiral fluctuations. The two states would be distinguished by the $q_y$ dependence of the susceptibility and very long chain lengths are required for these correlations to develop.  The interchain Neel susceptibility, $\chi_s$, together with finite-size scaling may provide one fruitful avenue for such studies.  For example, one could simply look for any indication of a sign change in $\chi_s$ with increasing $L$ for $J'/J<0.5$.  Our numerical studies are restricted to $L<30$, because of the high precision required to extract RG parameters and give no evidence of a sign change.   However, to within the expected accuracy, the numerics are consistent with the RG prediction that quadratic and cubic (in $J'/J$) fluctuation effects do not select any order.

%
%
%
%
%---------------------------------------------------------acknowledgments-----------------------------------------------------
\begin{acknowledgments}
The authors thank Leon Balents, Ganapathy Baskaran, John Berlinsky, Subhro Bhattacharjee, Oleg Starykh, and Andreas Weichselbaum for many useful discussions. This work was supported by NSERC and by the Shared Hierarchical Academic Research Computing Network. CK acknowledges support from the Canadian Institute for Advanced Research and Canada Research Chairs.

\end{acknowledgments}

%
%
%
%
%-----------------------------------------------------APPENDIX: Analytic Solution---------------------------------------------
\appendix*
\section{Approximate Analytic Solutions}
\label{appendix}

To better understand the flow of $g_{N}$ and $\gamma_{tw}$ under the RG, one may attempt solving a reduced subset of the given $\beta$-functions [Eqs.~(\ref{gbs}-\ref{ge})] analytically. If one drops terms in the $\beta$-functions that contribute at quartic and higher orders (in $J'/J$) to $g_{N}$, partial analytic solutions can be obtained that yield insight into the initial conditions corresponding to ``tuning'', i.e., to no exponential growth in $g_N$. This leaves us with the following reduced $\beta$-functions:

\begin{eqnarray}
	 \partial_{l}\gamma_{bs}&=&\gamma^{2}_{bs} \quad,\quad
  	 \partial_{l}\gamma_{M}=\gamma^{2}_{M} 
	 \\
	 \nonumber\\
	 \partial_{l}\gamma_{tw}&=&-\tfrac{1}{2}\gamma_{bs}\gamma_{tw}+\gamma_{M}\gamma_{tw}
	 \\
	 \nonumber\\
	 \partial_{l}g_{N}&=&g_{N}-\tfrac{1}{2}\gamma_{bs}g_{N}+\tfrac{1}{4}\gamma^{2}_{tw}\,,
\end{eqnarray}
which lead to the analytic expressions:
\begin{eqnarray}
	 \gamma_{bs}(l)&=&\frac{\gamma_{bs}(0)}{1-\gamma_{bs}(0)l} \quad,\quad
  	 \gamma_{M}(l)=\frac{\gamma_{M}(0)}{1-\gamma_{M}(0)l} 
	 \\
	 \nonumber\\
	 \gamma_{tw}(l)&=&\gamma_{tw}(0)\frac{\sqrt{1-\gamma_{bs}(0)l}}{1-\gamma_{M}(0)l}
	 \\
	 \nonumber\\
	 g_{N}(l)&=&
	\left[ 
      \frac{\gamma^{2}_{tw}(0)}{4}\int_{0}^{l}\frac{e^{-t}\sqrt{1-\gamma_{bs}(0)t}}{\left(1-\gamma_{M}(0)t\right)^{2}}\text{d}{t}
	  +g_{N}(0) 
	\right]
	\nonumber\\
	&&e^{l}\sqrt{1-\gamma_{bs}(0)l}
	\label{analytic_gN}
\end{eqnarray}

For a non-zero $\gamma_{M}(0)$, the integral in Eq.~(\ref{analytic_gN}) cannot be expressed in terms of elementary functions. However, assuming $\gamma_{M}(0)l\ll1$, it is possible to expand $\tfrac{1}{(1-\gamma_{M}t)^2}$ and find an expression for $g_{N}(l)$ as series of lower, incomplete gamma-functions:

\begin{eqnarray}
	g_{N}(l)
	&=&
	\Bigg[
	\frac{\gamma^{2}_{tw}(0)}{4}e^{-\tfrac{1}{\gamma_{bs}(0)}}\sqrt{-\gamma_{bs}(0)}\sum_{n}(n+1)\gamma^{n}_{M}F_{n}(l)
	\nonumber\\
	&&
	+g_{N}(0)
	\Bigg]e^{l}\sqrt{1-\gamma_{bs}(0)l}
	\label{analytic_gN_expansion}		
\end{eqnarray}
where $F_{n}(l)$ is given recursively by:

\begin{eqnarray}
	F_{n}(l)&=&
	\Bigg[
		\Gamma_{L}( n+\tfrac{3}{2}, l-\tfrac{1}{\gamma_{bs}(0)}) - 
		\Gamma_{L}( n+\tfrac{3}{2}, -\tfrac{1}{\gamma_{bs}(0)})
	\nonumber \\
	&& 
		- \sum_{m=1}^{n}\frac{(-1)^{m} n!}{\gamma^{m}_{bs}(0)(n-m)!} F_{n-m}(l)	
	\Bigg]
\end{eqnarray}

Each term in Eq.~(\ref{analytic_gN_expansion}) is accompanied by a power of $\gamma_{M}(0)$, and also we have a $\gamma^{2}_{tw}(0)$ factor that multiplies the sum. Since the above analytic expressions for the couplings were derived by neglecting quartic and higher-order contributions, we dismiss all terms in the expansion except the first two terms. Finally, using the above analytic expressions, the tuned initial condition for the Neel-Neel coupling, $g^{\text{\tiny crit}}_{N}(0)$, is found [$\gamma_{bs}(0)=-0.23$]:
\begin{equation}
	g^{\text{\tiny crit}}_{N}(0)=-\Big[0.27620(J'/J)^{2}+1.20152(J'/J)^{3}+\mathcal{O}(J'/J)^{4}\Big].
	\label{gncrit}
\end{equation}
It is also possible to determine the coefficient of the quartic term in the above expansion numerically, which gives $8.0$. Note that we have taken $g^{\text{\tiny crit}}_{N}(0)$ as the initial value for which $g_{N}(l_{\text{\tiny final}})=0$ [$\gamma_{tw}(l_{\text{\tiny final}})=1$] which for any $J'\lesssim 0.2$ is not any different from other ways of defining tuning (e.g. $g_{N}(l_{\text{\tiny final}})=-1$).  However,  the estimated value, $J'_c/J$, for the transition from CAF to spiral, would be smaller if $g_N(l_{\text{\tiny final}})<0$ was used. 

Finally, note that in the above calculation $\gamma_{bs}(0)=-0.23$ was used, which is the coupling of backscattering term at $4a_0$ rather than $a_0$ where $\gamma_{M}$ and $\gamma_{tw}$ have their bare values. Of course nothing is special about $a_0$ and one can repeat the same analysis at any arbitrary length, knowing the values of the couplings at that length scale. Thus what has been neglected is the small growth in $\gamma_{M}$ and $\gamma_{tw}$ from $a_0$ to $4a_0$. Taking into account this growth of these couplings, the quadratic and cubic parts of $g^{\text{\tiny crit}}_{N}(0)$ become, respectively, $0.291374$ and $1.31288$.
%
%
%
%____________________________________________________________________________________________________________________________
\bibliographystyle{apsrev4-1}
\bibliography{./Paper_arXiv_V3}

\end{document}